# Large-Signal Model of Graphene Field-Effect Transistors — Part I: Compact Modeling of GFET Intrinsic Capacitances

Francisco Pasadas and David Jiménez

*Abstract*—We present a circuit-compatible compact model of the intrinsic capacitances of graphene field-effect transistors (GFETs). Together with a compact drain current model, a large-signal model of GFETs is developed combining both models as a tool for simulating the electrical behavior of graphene-based integrated circuits, dealing with the DC, transient behavior, and frequency response of the circuit. The drain current model is based in a drift-diffusion mechanism for the carrier transport coupled with an appropriate field-effect approach. The intrinsic capacitance model consists of a 16-capacitance matrix including self-capacitances and transcapacitances of a four-terminal GFET. To guarantee charge conservation, a Ward-Dutton linear charge partition scheme has been used. The large-signal model has been implemented in Verilog-A, being compatible with conventional circuit simulators and serving as a starting point toward the complete GFET device model that could incorporate additional non-idealities.

*Index Terms*—Compact model, drift-diffusion, field-effect transistor, graphene, intrinsic capacitance, Verilog-A.

## I. Introduction

EXPERIMENTAL research into graphene field-effect transistors (GFETs) has rapidly increased in the past few years, mainly because of the potentially achievable high-speed performance [1]–[3]. However, there has been little exploration on the physical behavior of these devices under dynamic conditions.

Most circuits operate under time-varying terminal voltage excitation of the constituting devices. Depending on the magnitude of the time-varying signals, the dynamic operation can be classified as large-signal operation and small-signal operation. Both types of dynamic operation are influenced by the capacitive effects of the device, rendering indeed essential for eventual circuit design to derive reliable compact models encompassing such capacitive effects. Several intrinsic capacitance models for field-effect transistors (FETs) have been developed along the years. Basically, they can be categorized into two groups: 1) Meyer [4] and Meyer-like capacitance models and 2) charge-based capacitance models. The advantages and shortcomings of the two groups of models have been widely discussed and both of them have been implemented in circuit simulators [5], [6].

Although Meyer and Meyer-like models exhibit well-known problems with some circuits (e.g. DRAMs and switched capacitor filters), these compact models are widely used because of its simplicity and fast computation. But they assume that the capacitances in the intrinsic FET are reciprocal (as 2-terminal lumped capacitances), which is not the case in real devices, and earlier models based on this assumption cannot ensure charge conservation [7], [8]. Furthermore, most of the GFET capacitance models hitherto found in the literature are directly based upon such Meyer assumption and, therefore, may incorrectly interpret and predict the frequency performance of these devices. Examples of compact Meyer-like capacitance models of three-terminal devices based on drift-diffusion (DD) theory have been proposed by Rodríguez *et al*. [9], Zebrev *et al.* [10], Champlain [11], or Frégonèse *et al.* [12]. On the other hand, Habibpour *et al.* have proposed a semi-empirical large-signal GFET model based on a small set of fitting parameters, including the intrinsic capacitances $C_{gs}$, $C_{gd}$ and $C_{ds}$ which are extracted from S-parameters and dc measurements [13]. In this model the intrinsic capacitances are not bias dependent, so the model can be inaccurate depending on the selected bias point.

Charge-based models ensure charge conservation and consider the nonreciprocal property of capacitances in a FET. These features are required especially for radio-frequency (RF) applications in which the influence of transcapacitances are critical and should be considered. Thanks to some corrections assembled by Ward and Dutton [14] the charge-conservation issue was solved at the cost of introducing a capacitive-matrix which adds a bit of complexity.

Note that both Meyer and charge-based modeling approaches assume the so called *quasi-static-operation* approximation, where the fluctuation of the varying terminal voltages is assumed to be slow, so the stored charge could follow the voltages variations. Such an approximation is found

Manuscript received Month Day, Year; accepted Month Day, Year. Date of publication Month Day, Year; date of current version Month Day, Year. This work has received funding from the European Union's Horizon 2020 research and innovation programme under grant agreement No 696656 and from Ministerio de Economía y Competitividad under Grant TEC2012-31330 and Grant TEC2015-67462-C2-1-R.

The authors are with the Departament d'Enginyeria Electrònica, Escola d'Enginyeria, Universitat Autònoma de Barcelona, 08193 Bellaterra, Barcelona, Spain (e-mail: francisco.pasadas@uab.es, david.jimenez@uab.es).

This paper contains multimedia material available online at http://ieeexplore.ieee.org (File size: kB).

Color versions of one or more of the figures in this paper are available online at http://ieeexplore.ieee.org.

Digital Object Identifier 10.1109/TED.2016.2570426

to be valid when the transition time for the voltage to change is less than the transit time of the carriers from source to drain. This approximation works well in many FETs circuits, but it sometime fails, especially with long channel devices operating at high switching speeds, when the load capacitance is very small, and for digital circuits [5], [6].

Moreover, ambipolar electronics based on the symmetric I-V relation around the Dirac voltage is a key application of GFET technology. To take advantage of the ambipolarity it is essential: (1) controlling the device polarity and (2) tuning properly the Dirac voltage of a GFET in a circuit. The inclusion of a back-gate thus is essential for getting that tunability, which motivates the study of a general 4-terminal device. Examples are: (1) the polarity-controllable graphene inverter and voltage controlled resistor [15], [16]; and (2) the graphene-based frequency tripler [17] that has been demonstrated with a properly adjusted threshold voltage separation of two graphene FETs connected in series by a back-gate bias.

This is the first part of a two-part paper where we present a compact charge-based intrinsic capacitance model for double-gate four-terminal GFETs derived from a Ward – Dutton's linear charge partition scheme [14], which guarantees charge conservation. The model has been built from a field-effect model and DD carrier transport. We have developed explicit closed-form expressions for the 9 independent capacitances out of 16 capacitances in total, corresponding to 4 self-capacitances and 12 transcapacitances, covering continuously all the operation regions. Additionally, they have been written in Verilog-A, a language suited to circuit simulators.

In the second part of this paper [18], the large-signal model encompassing the drain current and the intrinsic capacitance models, both presented in the next section, will be assessed at the circuit level. A Verilog-A version of it is available online at http://ieeexplore.ieee.org. The DC bias point, the AC response, transient behavior and analysis of S-parameters have been compared to measurements from GFET based circuits that take advantage of ambipolar electronics.

## II. COMPACT CHARGE-BASED INTRINSIC CAPACITANCE MODEL OF GFETs

In this section, we provide a description of the charge-based intrinsic capacitance model of a four-terminal GFET. First of all, we investigate the device's electrostatics, which forms the basis to later formulate the drain current, which is based on a DD approach. Next, we have formulated appropriate models for the charge and capacitance, which are needed for any transient dynamics or frequency response simulation.

### A. Electrostatics of a four-terminal GFET

Let us consider a GFET with top and back gates, with the cross-section depicted in Fig. 1a. The electrostatics of the GFET can be understood using the equivalent circuit depicted in Fig. 2, which has been reported in [19]:

$$Q_{net}(x) = -C_t \left[ V_{gs} - V_{gs0} - V(x) + V_c(x) \right] \\ - C_b \left[ V_{bs} - V_{bs0} - V(x) + V_c(x) \right] \quad (1)$$

where $Q_{net} = q(p-n)$ is the overall net mobile sheet charge density where $q$ is the elementary charge, and $p$ and $n$ are the hole and electron carrier densities, respectively. $C_t = \varepsilon_0\varepsilon_t/L_t$ and $C_b = \varepsilon_0\varepsilon_b/L_b$ are the top and bottom oxide capacitances, respectively; $V_{gs}$-$V_{gs0}$ and $V_{bs}$-$V_{bs0}$ are the top and bottom gate voltage overdrive. These quantities comprise work-function differences between the gates and the graphene channel and possible additional charge due to impurities or doping; $V_c$ is the voltage drop across the graphene layer; $V(x)$ is the quasi-Fermi level along the graphene channel. This quantity must fulfill the boundary conditions: 1) $V(x=0) = 0$ at the source end; 2) $V(x=L) = V_{ds}$ (drain-source voltage) at the drain end.

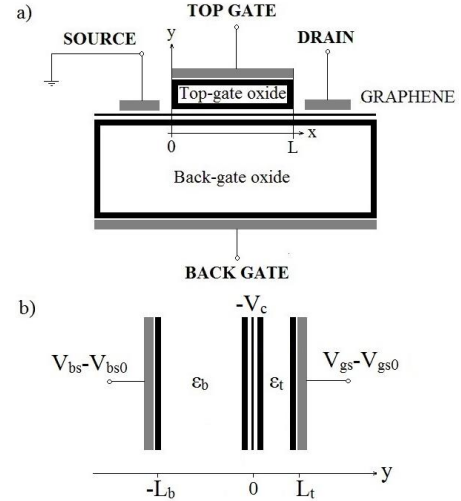

Fig. 1. a) Cross section of the GFET. A graphene sheet plays the role of the active channel. The electrostatic modulation of the carrier concentration in the 2D sheet is achieved via a double-gate stack consisting of top and bottom gate dielectrics and corresponding metal gates. The source is grounded and considered as the reference potential in the device. b) Scheme of the monolayer graphene based capacitor showing the relevant physical and electrical parameters, charges and potentials.

The energy $qV_c$ represents the shift of the Fermi level respect to the Dirac energy or, equivalently, $V_c$ represents, in the equivalent circuit, the voltage drop across the quantum capacitance $C_q$, which is pretty the same concept that the surface potential in conventional silicon transistors. This quantity is usually defined as $C_q = dQ_{net}/dV_c$ and has to do with the two-dimensional density of states of the monolayer graphene [20]. Fig. 3 shows a scheme of these potentials. The relation between $V_c$ and the quantum capacitance is given by $C_q = k|V_c|$, where $k = (2q^3/\pi\hbar^2v_F^2)$, and $v_F = (3a\gamma_0/2\hbar)$ is the Fermi velocity, where $\hbar$ is the reduced Planck's constant, a = 2.49 Å [21] is the carbon-carbon distance of honeycomb-like crystal lattice structure; and $\gamma_0$ = 3.16 eV [22] is the interlayer coupling. The $C_q$ expression is valid under the condition $qV_c \gg k_BT$, where $k_B$ is the Boltzmann constant and $T$ is the temperature; nevertheless, we have used this expression in order to keep the simplicity.

Applying circuit laws to the equivalent capacitive circuit, and noting that the overall net mobile sheet charge density in the graphene channel is equal to $Q_{net} = (1/2)C_qV_c$, the expressions (8) and (9) of $V_c$ are obtained, where the positive (negative) sign applies when $C_t(V_{gs}$-$V_{gs0}$-$V(x)) + C_b(V_{bs}$-$V_{bs0}$-$V(x)) < 0$ (>0).

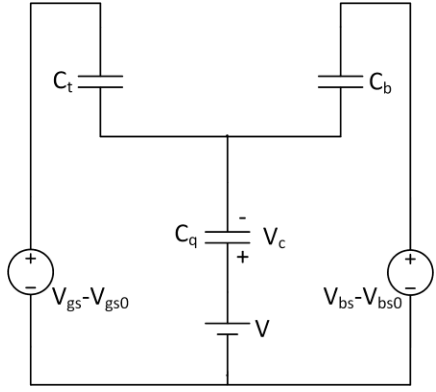

Fig. 2. Equivalent capacitive circuit of the GFET

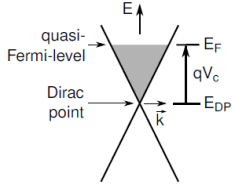

Fig. 3. Scheme of the energy-dispersion relation of graphene, showing the potential definitions employed in this work. $E_F$ is the quasi-Fermi-level energy; and $E_{DP}$ is the energy at the Dirac point (where the conduction band and the valence band touch each other).

### B. Drain Current

The drain current model presented in this Section is based on the DD theory, which is applicable while the transistor gate length is larger than the mean free path (MFP). The determination of the MFP in graphene is not trivial due to the strong dependence of the graphene sheet quality. Under practical conditions for common dielectric substrates, room temperature and ambient environment, MFPs of less than a hundred nm have been registered [23]. In most experiments reported up to now the prototype devices present channel lengths greater than the MFP, so we have considered carrier transport under a DD framework. The drain current can be written under the form $I_{ds} = -WQ_{tot}(x)v(x)$, where $W$ is the gate width, $Q_{tot}(x) = Q_t(x)+\sigma_{pud}$ is the free carrier sheet density along the channel at position $x$, $Q_t = q(p+n)$ is the transport sheet charge density, and $\sigma_{pud} = q\Delta^2/\pi\hbar^2 v_F^2$ is the residual charge density due to electron-hole puddles, with $\Delta$ being the inhomogeneity of the electrostatic potential [24]. Under the condition of symmetrical electron and hole mobilities [25], the total transport sheet charge density $Q_{tot}$ is expressed as a quadratic polynomial:

$$Q_{tot}(x) = \frac{q\pi(k_B T)^2}{3(\hbar v_F)^2} + \frac{q^3 V_c^2(x)}{\pi(\hbar v_F)^2} + \sigma_{pud} \quad (2)$$

In order to keep the model simplicity, the velocity saturation effect for the drift carrier velocity has not been considered, so $v(x) = \mu F(x)$, where $F(x) = -dV(x)/dx$ is the electric field along the channel and $\mu$ represents the effective carrier mobility, considered the same for both electrons and holes. Nevertheless, useful guidelines for including a simple velocity saturation model are given in the Appendix A.

The drain current equation must be integrated over the device length ($L$), and it is convenient to solve the integral using $V_c$ as the integration variable, consistently expressing $Q_{tot}$ as a function of $V_c$. All in all, we can write the drain current as:

$$I_{ds} = \mu \frac{W}{L} \int_{V_{cs}}^{V_{cd}} Q_{tot}(V_c) \frac{dV}{dV_c} dV_c \quad (3)$$

where $V_{cs}$ and $V_{cd}$ are obtained from (9), with $V_{cs} = V_c|_{V=0}$ and $V_{cd} = V_c|_{V=V_{ds}}$. In addition, the quantity $dV/dV_c$ can be derived from (8) and reads as follows:

$$\frac{dV}{dV_c} = 1 + \frac{C_q(V_c)}{C_t + C_b} \quad (4)$$

Useful explicit closed-form expressions for the intrinsic transconductance ($g_m = \partial I_{ds}/\partial V_{gs}$), back-gate transconductance ($g_{mb} = \partial I_{ds}/\partial V_{bs}$) and output conductance ($g_{ds} = \partial I_{ds}/\partial V_{ds}$) are given in the Appendix B.

### C. Charge Model

An accurate modeling of the intrinsic capacitances of FETs requires an analysis of the charge distribution in the channel versus the terminal bias voltages. So the terminal charges $Q_g$, $Q_b$, $Q_d$, and $Q_s$ associated with the top gate, bottom gate, drain, and source electrodes of a four-terminal device has been considered. For instance, $Q_g$ can be calculated by integrating $Q_{net\_g}(x) = C_t(V_{gs}-V_{gs0}-V(x)+V_c(x))$ along the channel and multiplying it by the channel width $W$. This expression for $Q_{net\_g}(x)$ has been obtained after applying Gauss' law to the top-gate stack shown in Fig. 1. A similar expression can be found for $Q_b$. It is worth noticing that:

$$Q_g + Q_b = -W \int_0^L Q_{net}(x) dx \quad (5)$$

On the other hand, the charge controlled by both the drain and source terminals can be computed based on Ward-Dutton's linear charge partition scheme [14], which guarantees charge conservation. The resulting equations are listed next:

$$Q_g = \frac{WC_t}{C_t + C_b}\left[ C_b L(V_{gs} - V_{gs0} - V_{bs} + V_{bs0}) - \int_0^L Q_{net}(x) dx \right]$$

$$Q_b = \frac{WC_b}{C_t + C_b}\left[ C_t L(V_{bs} - V_{bs0} - V_{gs} + V_{gs0}) - \int_0^L Q_{net}(x) dx \right] \quad (6)$$

$$Q_d = W \int_0^L \frac{x}{L} Q_{net}(x) dx$$

$$Q_s = -(Q_g + Q_b + Q_d)$$

The above expressions can conveniently be written using $V_c$ as the integration variable, as it was done to model the drain current. Based on the fact that the drain current is the same at any point $x$ in the channel (assuming there are not any generation-recombination processes involved), we get from the DD transport model the following equations, which are needed to evaluate the charges in (6):

$$dx = \frac{\mu W}{I_{ds}} Q_{tot}(V_c) \frac{dV}{dV_c} dV_c$$

$$x = \frac{\mu W}{I_{ds}}\left[ \int_{V_{cs}}^{V_c} Q_{tot}(V_c) \frac{dV}{dV_c} dV_c \right] \quad (7)$$

$$V_c(x) = -\frac{C_t}{C_t + C_b + \frac{1}{2}C_q}(V_{gs} - V_{gs0} - V(x)) - \frac{C_b}{C_t + C_b + \frac{1}{2}C_q}(V_{bs} - V_{bs0} - V(x)) \quad (8)$$

$$V_c(x) = \frac{(C_t + C_b) - \sqrt{(C_t + C_b)^2 \pm 2k\left[C_t(V_{gs} - V_{gs0} - V(x)) + C_b(V_{bs} - V_{bs0} - V(x))\right]}}{\pm k} \quad (9)$$

### D. Charge-based Capacitance Model

A four-terminal FET can be modeled with 4 self-capacitances and 12 intrinsic transcapacitances, which makes 16 capacitances in total. The capacitance matrix is formed by these capacitances where each element $C_{ij}$ describes the dependence of the charge at terminal $i$ with respect to a varying voltage applied to terminal $j$ assuming that the voltage at any other terminal remain constant.

$$C_{ij} = -\frac{\partial Q_i}{\partial V_j} \quad i \neq j \qquad C_{ij} = \frac{\partial Q_i}{\partial V_j} \quad i = j \quad (10)$$

where $i$ and $j$ stand for $g$, $d$, $s$, and $b$.

$$\begin{bmatrix} C_{gg} & -C_{gd} & -C_{gs} & -C_{gb} \\ -C_{dg} & C_{dd} & -C_{ds} & -C_{db} \\ -C_{sg} & -C_{sd} & C_{ss} & -C_{sb} \\ -C_{bg} & -C_{bd} & -C_{bs} & C_{bb} \end{bmatrix} \quad (11)$$

Each row must sum to zero for the matrix to be reference-independent, and each column must sum to zero for the device description to be charge-conservative. Note that of the 16 intrinsic capacitances only 9 are independent. Just to illustrate the procedure, $C_{dd}$ and $C_{db}$ have been calculated as,

$$C_{dd} = \frac{\partial Q_d}{\partial V_{cd}} \times \frac{\partial V_{cd}}{\partial V_d}$$

$$C_{db} = -\frac{\partial Q_d}{\partial V_{cd}} \times \frac{\partial V_{cd}}{\partial V_b} - \frac{\partial Q_d}{\partial V_{cs}} \times \frac{\partial V_{cs}}{\partial V_b} \quad (12)$$

where explicit closed-form expressions of the independent intrinsic capacitances have been implemented in Verilog-A to build the large-signal model. In the derivation of the capacitances, we have used:

$$\frac{\partial V_{cs}}{\partial V_g} = \frac{\partial V_{cd}}{\partial V_g} = \frac{C_t}{C_b}\frac{\partial V_{cs}}{\partial V_b} = \frac{C_t}{C_b}\frac{\partial V_{cd}}{\partial V_b} = -\frac{C_t}{C_t + C_b + C_q(V_c)}$$

$$\frac{\partial V_{cs}}{\partial V_s} = \frac{\partial V_{cd}}{\partial V_d} = \frac{C_t + C_b}{C_t + C_b + C_q(V_c)} \quad (13)$$

$$\frac{\partial V_{cs}}{\partial V_d} = \frac{\partial V_{cd}}{\partial V_s} = 0$$

Finally, from Eqs. (6) and (13) the following relations between the top and back gate capacitances can be worked out:

$$C_{bd} = C_{gd}\left(\frac{C_b}{C_t}\right) \qquad C_{db} = C_{dg}\left(\frac{C_b}{C_t}\right)$$

$$C_{bs} = C_{gs}\left(\frac{C_b}{C_t}\right) \qquad C_{sb} = C_{sg}\left(\frac{C_b}{C_t}\right) \quad (14)$$

$$C_{bb} = C_{gg}\left(\frac{C_b}{C_t}\right)^2 \qquad C_{bg} = C_{gb} = -C_{gg}\left(\frac{C_b}{C_t}\right) = -C_{bb}\left(\frac{C_t}{C_b}\right)$$

### III. INTRINSIC CAPACITANCE MODEL ASSESSMENT

In this section we present the intrinsic capacitances of a prototype GFET, which are derived from the model explained in the section before. Just to mention an example the GFET can be used as a key component for a frequency doubler [26]. To face the calculation of the transient behavior or frequency response of the circuit, it is essential to know how the intrinsic capacitances are related with the terminal voltages, which is exactly what the model we have presented do.

As for the prototype GFET we have used the one considered in [26] and described in Table I, which is a double-gated transistor with $C_t/C_b \approx 185$. A set of independent intrinsic capacitances have been plotted in Fig. 4 and Fig. 5 as a function of $V_{gs}$ and $V_{ds}$, respectively. A thorough discussion of the terminal charges and capacitances for the different operation regions can be found in [19] and [27], and could be directly applied to these results.

The accuracy of the developed compact intrinsic capacitance model around the Dirac point is benchmarked against a direct numerical solution of the problem using the MATLAB software [28]. In doing so, we have also implement a numeric solution of the drain, charge and capacitance models of the GFET as done in [27] but for the monolayer case and, therefore, using the exact solution of the derivative $C_q = dQ_{net}/dV_c$ for the quantum capacitance [29]:

$$C_q = \frac{2k_BT}{\pi(\hbar v_F)^2}\ln\left[2\left(1+\cosh\left(\frac{qV_c}{k_BT}\right)\right)\right] \quad (15)$$

All capacitances of the charge-based model resulted accurate around the Dirac point and the continuity has been also guaranteed.

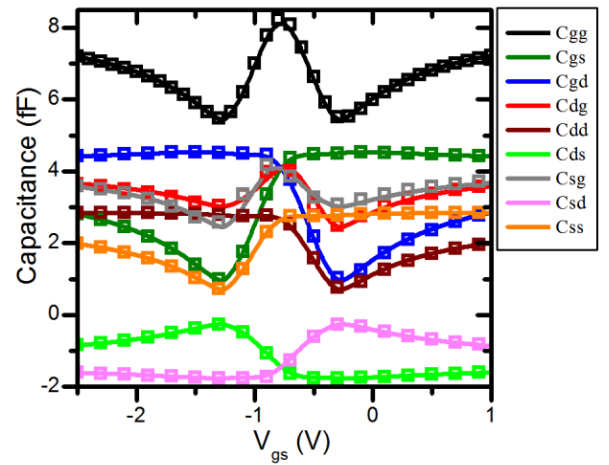

Fig. 4 (Color online) Compact model (solid lines) and numerical (symbols) calculation of the intrinsic capacitances versus the gate bias, assuming $V_{ds} = 1$ V for the device described in Table I.

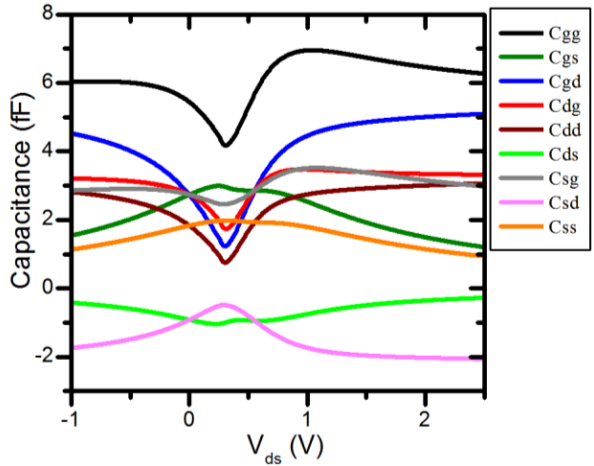

Fig. 5 (Color online) Intrinsic capacitances versus the drain bias for the device described in Table I. The calculations were done assuming $V_{gs}$ = -1 V.

The drain current model and the charge-based compact intrinsic capacitance description have been integrated in a circuit simulator, both written in Verilog-A, which is a standard language used in circuit simulators. The complete large-signal model is available online at http://ieeexplore.ieee.org. The intrinsic large-signal GFET equivalent circuit is depicted in Fig. 6.

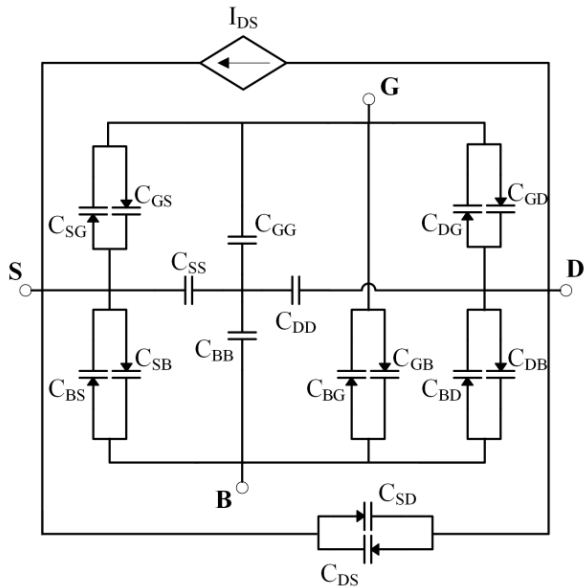

Fig. 6 Large-signal GFET equivalent circuit formed by the drain current model and the intrinsic capacitance model.

TABLE I. INPUT PARAMETERS OF THE GFET UNDER TEST.

| Input parameter | Value | Input parameter | Value |
| --- | --- | --- | --- |
| $T$ | 300 K | $W$ | 0.84 µm |
| $\mu$ | 1300 cm$^2$/Vs | $L_t$ | 5 nm |
| $V_{gs0}$ | -1.062 V | $L_b$ | 300 nm |
| $V_{bs0}$ | 0 V | $\varepsilon_{top}$ | 12 |
| $\Delta$ | 0.140 eV | $\varepsilon_{bottom}$ | 3.9 |
| $L$ | 0.5 µm | | |

## IV. CONCLUSIONS AND FUTURE PROSPECTS

In conclusion, we have presented a compact large-signal model of GFETs suitable for circuit level design. A drain current model and a charge-based intrinsic capacitance model have been proposed assuming a field-effect model and drift-diffusion carrier transport. The model can predict the bias dependence of small-signal parameters at HF operation and it correctly describes the nonlinear behavior of the device allowing for the simulation of intermodulation distortion and high frequency large-signal operation. The model is physics-based so that it can be used as a predictive tool for graphene-based RF applications.

The intrinsic capacitance model proposed here is a starting point toward a complete GFET device model incorporating additional device non-idealities. On the one hand, a saturation model for the carrier velocity has to be included, consistently with the numerical studies of electronic transport in monolayer graphene relying on Monte Carlo simulations [30]. Moreover, it has been realized that an accurate and physical description of a mobility is essential for distortion analysis [6]. Further inclusions of many important physical effects such as short-channel and narrow width effects, trapped charge, channel-length modulation, non-uniform doping effect, and so on, could also be important. On the other hand, the model has to correctly predict the HF noise, which is important for the design of, for example, low noise amplifiers. The model should also include the non-quasi static (NQS) effect, so it can properly describe the device behavior at very high-frequency where the quasi-static assumption could break down. The model presented applies to the intrinsic device, but an appropriate model of the device's parasitics has to be developed. A common modeling approach for RF applications is to build subcircuits based on the intrinsic FET, thus the parasitic elements must be included using simple subcircuits that also reduce the simulation time and make parameter extraction easier. These subcircuits should also be linked to process and geometry information to guarantee scalability and prediction capabilities of the model.

## APPENDIX A

A soft saturation model is usually considered for the drift carrier velocity in graphene as $v(x) = \mu F(x)/(1+\mu|F(x)|/v_{sat})$, where $v_{sat}$ is the saturation velocity and could be considered close to the Dirac point as a constant, and, for higher $V_c$ it has been found to follow the relation $|V_c|^{-1}$ [31], [32]. To include the velocity-saturation effects to the large-signal model, Eqs. (3) and (7) must be respectively substituted for:

$$I_{ds} = \mu \frac{W}{L + \mu \int_{V_{cs}}^{V_{cd}} \frac{1}{v_{sat}(V_c)} \frac{dV}{dV_c} dV_c} \int_{V_{cs}}^{V_{cd}} Q_{tot}(V_c) \frac{dV}{dV_c} dV_c \quad (A1)$$

$$dx = \frac{\mu W}{I_{ds}} Q_{tot}(V_c) \frac{dV}{dV_c} dV_c - \mu \frac{1}{v_{sat}(V_c)} \frac{dV}{dV_c} |dV_c|$$

$$x = \frac{\mu W}{I_{ds}} \left[ \int_{V_{cs}}^{V_c} Q_{tot}(V_c) \frac{dV}{dV_c} dV_c \right] - \mu \left| \int_{V_{cs}}^{V_c} \frac{1}{v_{sat}(V_c)} \frac{dV}{dV_c} dV_c \right| \quad (A2)$$

where $v_{sat}(V_c) = (2/\pi)v_F$ in case of assumption of a constant saturation velocity; or $v_{sat}(V_c) = |(2/\pi)v_F(\hbar\Omega/-qV_c)|$ for an energy-dependent velocity saturation, where $\hbar\Omega$ is the effective energy at which a substrate optical phonon is emitted [32].

APPENDIX B

In this section, useful closed-form expression for intrinsic $g_m$, $g_{mb}$ and $g_{ds}$ are provided as follows:

$$g_m = \frac{\partial I_{ds}}{\partial V_{gs}} = \mu \frac{W}{L} \frac{k}{2} \frac{C_t}{C_t + C_b}\left(V_{cs}^2 - V_{cd}^2\right) \quad (B1)$$

$$g_{mb} = \frac{\partial I_{ds}}{\partial V_{bs}} = \mu \frac{W}{L} \frac{k}{2} \frac{C_b}{C_t + C_b}\left(V_{cs}^2 - V_{cd}^2\right) \quad (B2)$$

$$g_{ds} = \frac{\partial I_{ds}}{\partial V_{ds}} = \mu \frac{W}{L} \frac{k}{6}\left[3V_{cd}^2 + (\pi k_B T)^2 + \frac{6\sigma_{pud}}{k}\right] \quad (B3)$$

Equation (B3) suggests an interesting discussion about the minimum output conductance. Figures of merit like the maximum frequency of oscillation (defined as the highest possible frequency where the magnitude of the power gain of the transistor is reduced to unity) is key for RF applications. Although experimental cut-off frequencies up to 427 GHz [33] have been achieved, only a maximum frequency of oscillation of 70 GHz [34] have been demonstrated. That maximum oscillation frequency is considered still very low [35]. In particular, the absence of a band gap in graphene prevents proper current saturation, so there is a lot of ongoing research trying to minimize $g_{ds}$. The expression (B3) above can be used to predict the minimum intrinsic output conductance at room temperature:

$$g_{ds}^{min}(T=300K) = \mu \frac{W}{L}(2.415 \cdot 10^{-4} \text{C/m}^2 + \sigma_{pud}) \quad (B4)$$

ACKNOWLEDGMENTS

This work has received funding from the European Union's Horizon 2020 research and innovation programme under grant agreement No 696656 and from Ministerio de Economía y Competitividad under Grant TEC2012-31330 and Grant TEC2015-67462-C2-1-R.